\RequirePackage{fix-cm}
\documentclass[twocolumn]{svjour3}          
\smartqed  
\usepackage{graphicx}
\begin{document}

\title{Linear and nonlinear optical properties of multi-layered spherical nano-systems with donor
impurity in the center}

\author{A. R. Jafari \and Y. Naimi}


\institute{A. R. Jafari
           \and
          Y. Naimi \at
              Department of Physics, Lamerd Branch, Islamic Azad University, Lamerd, Iran.\\
              \email{y.naimi@iaulamerd.ac.ir}.
}

\date{Received: date / Accepted: date}

\maketitle
\begin{abstract}
In this study, the linear, third-order nonlinear and total absorption coefficients (ACs)
of multi-layered quantum dot (MLQD) and  multi-layered quantum anti-dot (MLQAD)
with a hydrogenic impurity are calculated.
The analytical and numerical solutions of Schr\"{o}dinger equation for both MLQD and MLQAD,
within the effective mass approximation and dielectric continuum model, are obtained.
As our numerical results indicate, an increase in the optical intensity
changes the total AC considerably, but the intensity range that leads to
these changes is different for MLQAD and MLQD. It is observed that by changing
the incident photon energy, the AC curves corresponding to MLQAD and MLQD are
of different shapes and behaviors.  The peak heights of AC curves corresponding
to MLQAD are strongly affected by changing the core antidot radius and the shell
thickness values, however in these cases no considerable changes are observed
in peak heights of MLQD. Furthermore, in contrast
to MLQAD, the photon energies corresponding to total AC peaks
 of MLQD are more affected by changing the confining potentials (CPs).

\end{abstract}\maketitle
\noindent{keywords: Multi-layered quantum dots and anti-dots; Optical intensity; Linear and nonlinear absorbtion coefficients; Confining potential.

\section{Introduction}
In the recent years, nano-structured semiconductors have
attracted much attention, due to a wide range of their applications
in electronic and optoelectronic devices. The most known spherical nano-systems are single-layered quantum dots (QDs), multi-layered QDs
 and quantum anti-dots (QADs). Such systems lead to foundation of discrete and not degenerate energy levels like atoms
 \cite{davat,holo,varshini,sadeghi,vasegh,cheng,naimi}. The amount of energy is dependent on the material, the shape, the size,
 the confinement potential and the surrounding matrix of QDs. The nonlinear optical properties associated
with optical absorption of mentioned nano-structures are known to be much stronger than bulk material, due to the quantum confinement
effect \cite{atan,tsan,karimi,yang,xie,bas}. These particular properties and precise engineering enable us to construct the optoelectronic
devices, such as infrared photo detectors and high speed electro-optical modulators \cite{tro,jia,liu,len}. Due to relevance of QDs to several
technological applications, their linear and nonlinear optical properties have been investigated both theoretically and experimentally by many
authors \cite{saf,ru,li,zha,lia,kara,oz1,oz2,kho}. \\

The optical properties of QDs was considered by Efros \cite{efo} for the first time. He studied the light absorption coefficient in a
spherical QD with infinitely height walls. In ref \cite{saf}, the authors have calculated not only the linear and nonlinear absorption
coefficients but also the refractive index changes in a three-dimensional Cartesian coordinate quantum box with finite confining potential
barrier height. Ruihao Wei and Wenfang Xie in \cite{ru} have used the effective mass approximation and the perturbation theory for studying
the linear and nonlinear optical properties of a hydrogenic impurity confined in a disk-shaped QD with a parabolic potential in the presence
 of an electric field. Their results show that the optical properties are strongly affected by a confining strength and applied electric field
 intensity. In ref \cite{zha}, the nonlinear optical properties and the refractive index changes in a two-dimension (electron-hole system)
 have been investigated theoretically. The linear, nonlinear and total refractive index changes and absorption coefficients for a transitions
 in a spherical QD with parabolic potential have recently been studied in ref \cite{oz1}. Their result, expressed in several allowed
 transitions $1s-1p$, $1p-1d$ and $1d-1f$, show that  the transition between orbitals with big $l$ (orbital quantum number) values move to lower
 incident photon energy region in the presence of parabolic potential term. The effect of constant effective mass and position-dependent
 effective mass on the optical linear and nonlinear absorbtion coefficient have been calculated by Khordad in ref \cite{kho}.\\

For many semiconductor quantum heterostructures, such
as GaAs/Ga$_{1-x}$Al$_{x}$As, the polarization and image charge effects
can be significant in the multi-layered system if there
is a large dielectric discontinuity between the dot and the
surrounding medium. However, this is not the case for the
GaAs/Ga$_{1-x}$Al$_{x}$As quantum system \cite{Ada}, therefore these effects
can be ignored safely in our calculation. Furthermore, for the sake of generality, the difference between the electronic effective mass
in the dot (antidot) core, the shell and the bulk materials have been ignored (i.e., $m_{1}=m_{2}=m_{3}=m$ ). The energies are measured in
$meV$, the effective Rydberg, $Ry=me^4/2\hbar^2(4\pi\varepsilon)^2$, and distances are expressed in $a_{0}=4\pi\varepsilon\hbar^2/(me^2)$,
for instance, in the particular case of GaAs-based semiconductors, $m = 0.067me$ , and $\varepsilon = 13.18\varepsilon_{0}$.
Thus, for a GaAs host, the effective Rydberg is numerically $Ry = 5.2 meV$, and the effective Bohr radius,  $a_{0} = 10.4 nm$. \\
The rest of this paper contains the following sections. In sec 2 we solve the Schr\"{o}dinger equation for both MLQAD and MLQD with a
hydrogenic impurity in the center. The solutions in each case have been expressed in the Whittaker and hyper geometrical functions. In sec 3 the
linear, nonlinear and total ACs are plotted for various conditions as the function of incident photon energy. Finally, our conclusions are
presented in the last section.\\

\section {Theory and formulation}
The multi-layered spherical nano-systems can be made by binding of GaAs, Ga$_{1-x}$Al$_{x}$As and Ga$_{1-y}$Al$_{y}$As materials. The type
of these adjacent material connections next to each other leads to the so-called MLQD and MLQAD as shown in Fig.~\ref{f1}. More precisely, a MLQD (MLQAD)
consists of a spherical core made of GaAs (Ga$_{1-y}$Al$_{y}$As) surrounded by a spherical shell
of Ga$_{1-x}$Al$_{x}$As (Ga$_{1-x}$Al$_{x}$As), embedded in the bulk of Ga$_{1-y}$Al$_{y}$As (GaAs). In other words, the MLQAD is made whenever the core material and the bulk material in MLQD change places. We denote $a$ for the core radius
and $b$ for the total dot
(core plus shell) radius; therefore, $b-a$ is the thickness of the shell.
\begin{figure*}
\centering
\includegraphics[ height=6.18cm,width=15.7cm]{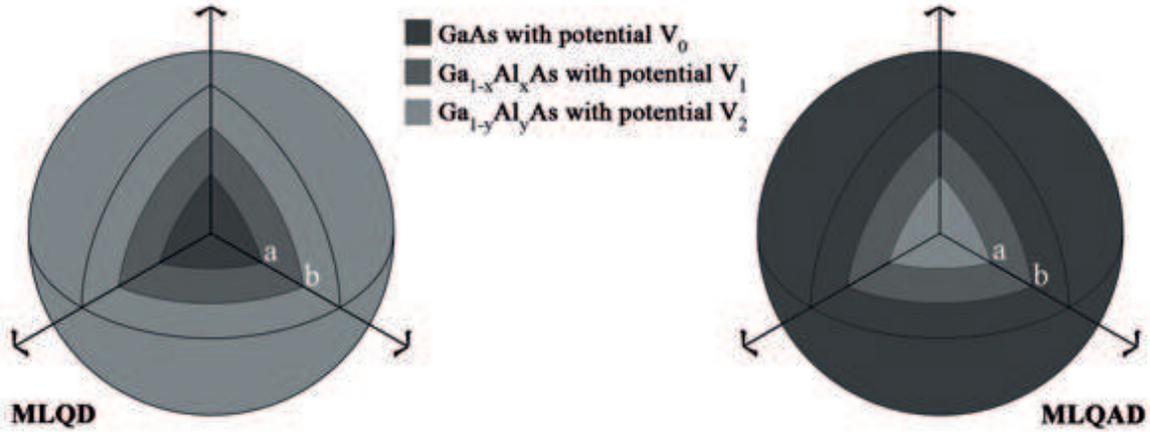}
\caption{A schematic view of the MLQD and MLQAD structures} \label{f1}
\end{figure*}

Within the framework of the effective mass approximation and in the spherical coordinate system, the Hamiltonian for the MLQD (MLQAD) with
 hydrogenic impurity in center is given by
\begin{equation}\label{1} \hat{H} =-\frac{\hbar^{2}}{2m(r)}\nabla^{2} -\frac{e^2}{4\pi\varepsilon r}+V(r)\end{equation}
where $ m(r)$ and $\varepsilon$ are the electronic effective mass and dielectric constant in the semiconductor medium.  $m(r)$ is represented
 as follows

 \begin{equation} \label{2} m(r) =  \left\{
 \begin{array}{ll}
   m_{1} & \hspace{1.5cm}r<a \\
   m_{2} &  \hspace{1.5cm} a\leq r \leq b\\
  m_{3} & \hspace{1.5cm} r>b
 \end{array}\right.
 \end{equation}
$V(r)$ is the CP and  arises due to a mismatch between the electronic
affinities of the two regions. The barrier height $V_{1}$ ($V_{2}$) arises in layer Ga$_{1-x}$Al$_{x}$As (Ga$_{1-y}$Al$_{y}$As) when it binds to the GaAs (Ga$_{1-x}$Al$_{x}$As). According to fig.~\ref{f1}, the potential energy $V(r)$ in Eq.~\ref{1}, for MLQD is
\begin{equation} \label{2} V(r) =  \left\{
\begin{array}{ll}
V_{0} & \hspace{1.5cm}r<a \\
V_{1} & \hspace{1.5cm} a\leq r \leq b\\
V_{2} & \hspace{1.5cm} r>b
\end{array}\right.
\end{equation}
and for MLQAD is
\begin{equation} \label{2} V(r) =  \left\{
\begin{array}{ll}
V_{2} & \hspace{1.5cm}r<a \\
V_{1} & \hspace{1.5cm} a\leq r \leq b\\
V_{0} & \hspace{1.5cm} r>b
\end{array}\right.
\end{equation}
the value of $V_{0}$ is always zero (i.e., the GaAs material have no CP) but writing $V_{0}$
helps us to be able to write next equations in the compact form as you will see.

Since the potential is spherically symmetric, we are going to use the spherical coordinates. In the three dimensional spherical coordinate,
 the Schr\"{o}dinger eigenvalue equation is $\hat{H} \psi (r,\Omega)=E\psi (r,\Omega)$, where the eigenfunctions are of
the separable form $\psi(r,\Omega)= R(r)Y(\Omega)$. For the $r$-dependent (spherically symmetric) potentials, the angular dependence of the
 wave function, $Y(\Omega)$, is given by the spherical harmonics \cite{zet}. In particular, the angular part is unaffected by the core and
 total radius, thus we ignore this part.
The radial part is affected by not only the core and total radius but also the dot or antidot CPs, and hence, we consider it in the following subsections.

\subsection{MLQAD}
Using the method of separation of variables in Schr\"{o}dinger equation (1), the radial eigenvalue equation in the spherical coordinate in
this case can be expressed as
 \begin{eqnarray}\label{3}
 &&\hspace{-0.5cm}\frac{\hbar^{2}}{2m}(\frac{d^{2}}{dr^{2}}+\frac{2}{r}\frac{d}{dr}-\frac{l(l+1)}{r^{2}})R^{(i)}(r)\nonumber \\
&+&(E-V_{i}+\frac{e^{2}}{4 \pi \varepsilon r})R^{(i)}(r)=0,
 \end{eqnarray}
where $i=2$, $1$ and $0$ are corresponding to $r<a$, $a\leq r\leq b$ and $r>b$ respectively. Just as we can see in \cite{naimi}, we can
solve this equation as follows. Using the following  definitions
\begin{equation}\label{6}\alpha^{2}_{i}=\frac{-8m (E-V_{i})}{\hbar^{2}},\ \ \ \ \xi_{i}=\alpha_{i}r,\ \ \ \lambda_{i}=\frac{2m e^{2}}{4\pi\varepsilon\hbar^{2}\alpha_{i}^2}.\end{equation}
and the change of variable as $W(\xi_{i})=\xi_{i} R^{(i)}(r)$, then the above Schr\"{o}dinger equation can be rewritten as following Whittaker
equation
 \vspace{0.3cm}
\begin{equation}\label{9}
\frac{d^{2}W(\xi_{i})}{d\xi_i^{2}}+\left(\frac{-1}{4}+\frac{\lambda_{i}}{\xi_{i}}+\frac{\frac{1}{4}-(l+\frac{1}{2})^{2}}{\xi^{2}_{i}}\right)
W(\xi_{i})=0,
\end{equation}
The general solution to this equation is the Whittaker functions \cite{abra}
 \vspace{0.3cm}
\begin{equation}\label{12} W(\xi_{i})=C_{1i}M_{\lambda_{i},l+\frac{1}{2}}(\xi_{i})+
C_{2i}U_{\lambda_{i}, l+\frac{1}{2}}(\xi_{i}),\end{equation}
where $C_{1i}$ and $C_{2i}$ are constants to be determined by continuity, asymptotic and normalization conditions.
The asymptotic behavior indicates  that the wave function must be finite and zero in origin neighborhood and far away
from origin, respectively. These conditions lead to $C_{22}=C_{10}=0$ (since $U_{\lambda_{i}, l+\frac{1}{2}}(\xi_{i})$ and $M_{\lambda_{i},l+\frac{1}{2}}(\xi_{i}$)
diverge at the origin and infinity, respectively). Now one can write the wave functions in three regions as
 \vspace{0.3cm}
 \begin{eqnarray}
 &&\hspace{-0.2cm}R^{(2)}=C_{12}\frac{M_{\lambda_{2},l+\frac{1}{2}}(\xi_{2})}{\xi_{2}},\ \ \ \hspace{2cm}r<a, \\
  &&\hspace{-0.2cm}R^{(1)}=\frac{C_{11}M_{\lambda_{1},l+\frac{1}{2}}(\xi_{1})+ C_{21}U_{\lambda_{1},l+\frac{1}{2}}(\xi_{1})}{\xi_{1}},
 \ \ \ \ a\leq r \leq b, \nonumber \\
 && \\&&\hspace{-0.2cm}R^{(0)}=C_{20}\frac{U_{\lambda_{0},l+\frac{1}{2}}(\xi_{0})}{\xi_{0}},\ \ \ \hspace{2cm}r>b.
  \end{eqnarray}

\subsection{MLQD}
The radial eigenvalue equation (5) is also valid for MLQD but according to different definitions of $V(r)$ for MLQD
(see Eq. (3) and (4)), in this case we should use $i=0$, $1$ and $2$ for $r<a$, $a\leq r\leq b$ and $r>b$ regions,
respectively. In the case of MLQD, the Eq. (5) has been solved in detail by Hsieh et al. in \cite{cheng}, so we do not
directly solve this equation and we shortly review their results for three different regions as follows
\begin{description}
  \item[(i=0)] Inside the core of system ($r<a$) with $V_{0}=0$, the positive energy bound
states are possible for an electron that is confined inside the core dot; therefore, solutions
of the  Schr\"{o}dinger equation can be studied in three energy states \cite{cheng}
\begin{itemize}
 \item[(A)] $E<0$
\vspace{0.3cm}
\begin{eqnarray}\label{3}
 &&\hspace{-0.5cm}R^{(0)}({\xi_{0A}}) \nonumber \\
 \hspace{-0.5cm}&=&C_{0A}e^{-\xi_{0A}/2}\xi_{0A}^{l}M(l+1-\lambda_{0},2l+2,\xi_{0A})
\end{eqnarray}
where $C_{0A}$ is the normalization constant and $M$ is the first kind of confluent hypergeometric function. The used parameters in the above solution are
\vspace{0.3cm}
\begin{equation}\label{8}
\xi_{0A}=\sqrt{\frac{-8mE}{\hbar^2}} r,\ \ \ \lambda_{0}=\frac{e^2}{4\pi\varepsilon\hbar}\sqrt{\frac{m}{-2E}}.
\end{equation}\
\item [(B)] $E>0$
\begin{equation}\label{3}
R^{(0)}(\xi_{0B})=C_{0B}\sum^{\infty}_{k=l+1}A^{l}_{k}(\beta_{0})\xi^{k-1}_{0B}.
\end{equation}
Similar to the case of $E<0$, $C_{0B}$ is the normalization constant. The recurrence
relation $A^{l}_{k}$ function and the other parameters are defined as
\vspace{0.3cm}
\begin{equation}\label{3}
A^{l}_{l+1}(\beta_{0})=1,
\end{equation}
\begin{equation}\label{3}
A^{l}_{l+2}(\beta_{0})=\frac{\beta_{0}}{l+1},
\end{equation}
\begin{eqnarray}\label{3}
&&\hspace{-0.4cm}A^{l}_{k}(\beta_{0})\nonumber \\
&=&\frac{2\beta_{0}A^{l}_{k-1}(\beta_{0})-A^{l}_{k-2}(\beta_{0})}{(k+l)(k-l-1)} \hspace{0.7cm} k>l+2,
\end{eqnarray}

\begin{equation}\label{8}
\xi_{0B}=\sqrt{\frac{8mE}{\hbar^2}} r,\ \ \ \ \beta_{0}=\frac{e^2}{4\pi\varepsilon\hbar}\sqrt{\frac{m}{2E}}.
\end{equation}
 \item [(C)] $E=0$
 \vspace{0.3cm}
\begin{equation}\label{3}
R^{(0)}(\alpha_{0C}r)=C_{0C}r^{(-1/2)} J_{2l+1}(\alpha_{0C}\sqrt{r}),
\end{equation}
\vspace{0.2cm}
$C_{0C}$ is the normalization constant and $J_{2l+1}$ is the Bessel function. $\alpha_{0C}$ is defined as
\vspace{0.3cm}
\begin{equation}\label{3}
\alpha_{0C}=\sqrt{\frac{8me^2}{4\pi\varepsilon\hbar^2}}.                                                                                             \end{equation}
            \end{itemize}
\vspace{0.3cm}
\item[(i=1)] For $a\leq r\leq b$ \\
Using the convenient parameters (6), the Schr\"{o}dinger equation in this region can be expressed in term of Whittaker equation as Eq. (7)
 which has a result just as the result of (10). In other words, the middle layer in both MLQD and MLQAD behaves in a similar way.
  \item[(i=2)] For $r>b$ \\
In this case, similar to the above case, the Schr\"{o}dinger equation can be expressed in terms of Whittaker equation
as Eq. (7) but by using the asymptotic condition in infinity (the M function diverges at infinity), the solution of equation can be written as
\begin{equation}\label{3}
R^{(2)}(\xi_{2})=C_{2}e^{-\xi_{2}/2}\xi^{l}_{2}U(l+1-\lambda_{2},2l+2,\xi_{2})
\end{equation}
where the parameters $\xi_{2}$ and $\lambda_{2}$ are defined as (6). For reading with full detail refer to \cite{cheng}.
\end{description}

\subsection{Linear and nonlinear absorption coefficients}
By using the density matrix formalism, the optical
properties for a spherical nano-system can be calculated. For
this purpose the system under study can be excited by
an external electromagnetic field of frequency $\omega$,
\begin{equation}\label{3}
  E(t)=\tilde{E}e^{i\omega t}+\tilde{E}e^{-i\omega t}
\end{equation}
If $E$ is the perpendicular electromagnetic field
along the $z$ axis, the Hamiltonian of system in this case is $H_{0}+ezE(t)$ where where $H_{0}$ is the Hamiltonian of system without
the electromagnetic field $E(t)$. The linear and third-order nonlinear optical
absorption coefficients of a spherical nano-system, within a two-level
system (in a special case, from ground state to first allowed excited state), can be expressed as \cite{saf,zha,kho}
\begin{equation}
\alpha^{(1)}(\omega)=\omega \sqrt{\frac{\mu}{\varepsilon_{R}}}\frac{\sigma_{v}\hbar\Gamma_{21}|M_{21}|^2}{(E_{21}-\hbar\omega)^2+(\hbar\Gamma_{21})^2},
\end{equation}
\begin{eqnarray}
&&\hspace{-0.5cm}\alpha^{(3)}(\omega, I)\nonumber \\
 &=&- \omega\sqrt{\frac{\mu}{\varepsilon_{R}}}(\frac{I}{2\varepsilon_{0}n_{r}c}) \frac{4\sigma_{v}\hbar\Gamma_{21}|M_{21}|^4}{[(E_{21}-\hbar\omega)^2+(\hbar\Gamma_{21})^2]^{2}}
\end{eqnarray}
where, $\mu$, $\sigma_{v}$ and $n_{r}$ are the permeability,  carrier density and refractive index  of the system respectively.
 $\hbar\omega$ is the incident photon energy, $\Gamma_{21}=1/T_{21}$, $T_{21}$ is the relaxing time between states 1 and 2, $I$
 is the optical intensity of incident wave and $c$ is the speed of light in the free space. The remained quantities are defined as
\begin{eqnarray}
&&E_{21}=E_{2}-E_{1},\nonumber \\
&&\varepsilon_{R}=n^{2}_{r}\varepsilon_{0}, \nonumber \\
&&M_{21}=|<\psi_{2}|ez|\psi_{1}>|,
\end{eqnarray}
where $M_{21}$ is an element of electric dipole moment matrix that in the spherical coordinate is $|<\psi_{2}|er\cos\theta|\psi_{1}>|$.

Finally, the total absorption coefficient is
\begin{equation}\label{27}
 \alpha(\omega,I) =\alpha^{(1)}(\omega)+\alpha^{(3)}(\omega,I).
\end{equation}

\section{Results and discussion}

In this section, we calculate the linear, nonlinear and total absorption coefficients for different shell thicknesses and CPs
for both MLQAD and MLQD. The unchanged parameters used in our calculations are : $\Gamma_{21}=0.2ps$, $\sigma_{v}=3.0\times10^{22}$, $n_{r}=3.2$, $\varepsilon_{R}=13.18$. \\
Figs.~\ref{f2}(a) and (b) are related to the linear, nonlinear and total absorption coefficients  as a function of incident photon
energy for MLQAD and MLQD respectively, when $I=20MW/m^{2}$, $a=2a_{0}$, $b-a=0.5a_{0}$, $V_{1}=2Ry$ and $V_{2}=5Ry$. From Fig.~\ref{f2},
 it is clear that the ACs of the MLQD and MLQAD have  different performances with respect to incident photon energy in the same condition.
  In the case of comparing the total AC curves,  Fig.~\ref{f2} shows that the MLQD has an approximately symmetric curve but MLQAD has
  a curve that abruptly increases then asymptotically goes to zero.  In the case of the nonlinear AC curves, it is obvious that for this
  selected intensity value, the MLQD has no significant nonlinear AC but the MLQAD shows a significant nonlinear AC.
\begin{figure*}
\mbox{\includegraphics[ height=2.7in,width=3.5in]{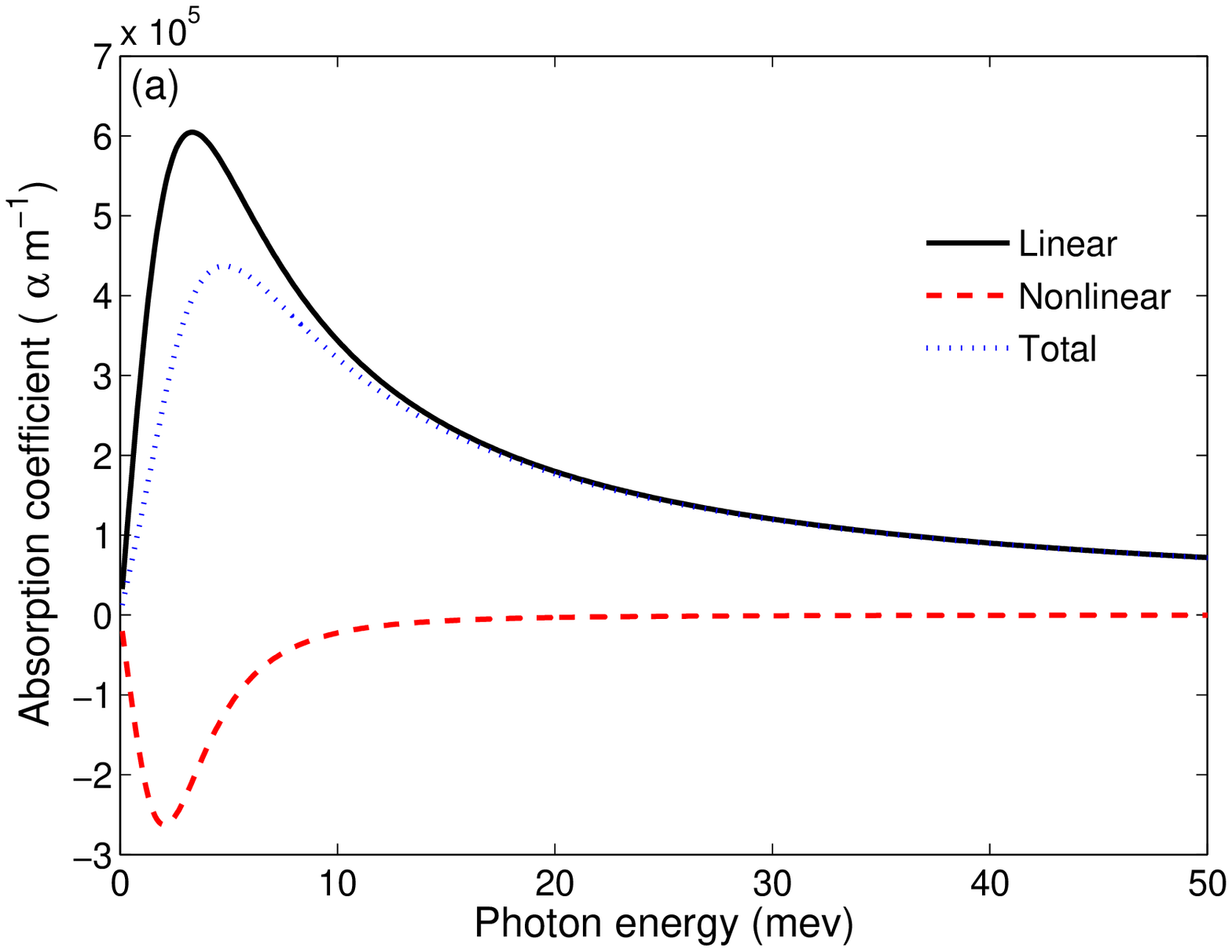}\quad
\includegraphics[ height=2.7in,width=3.5in]{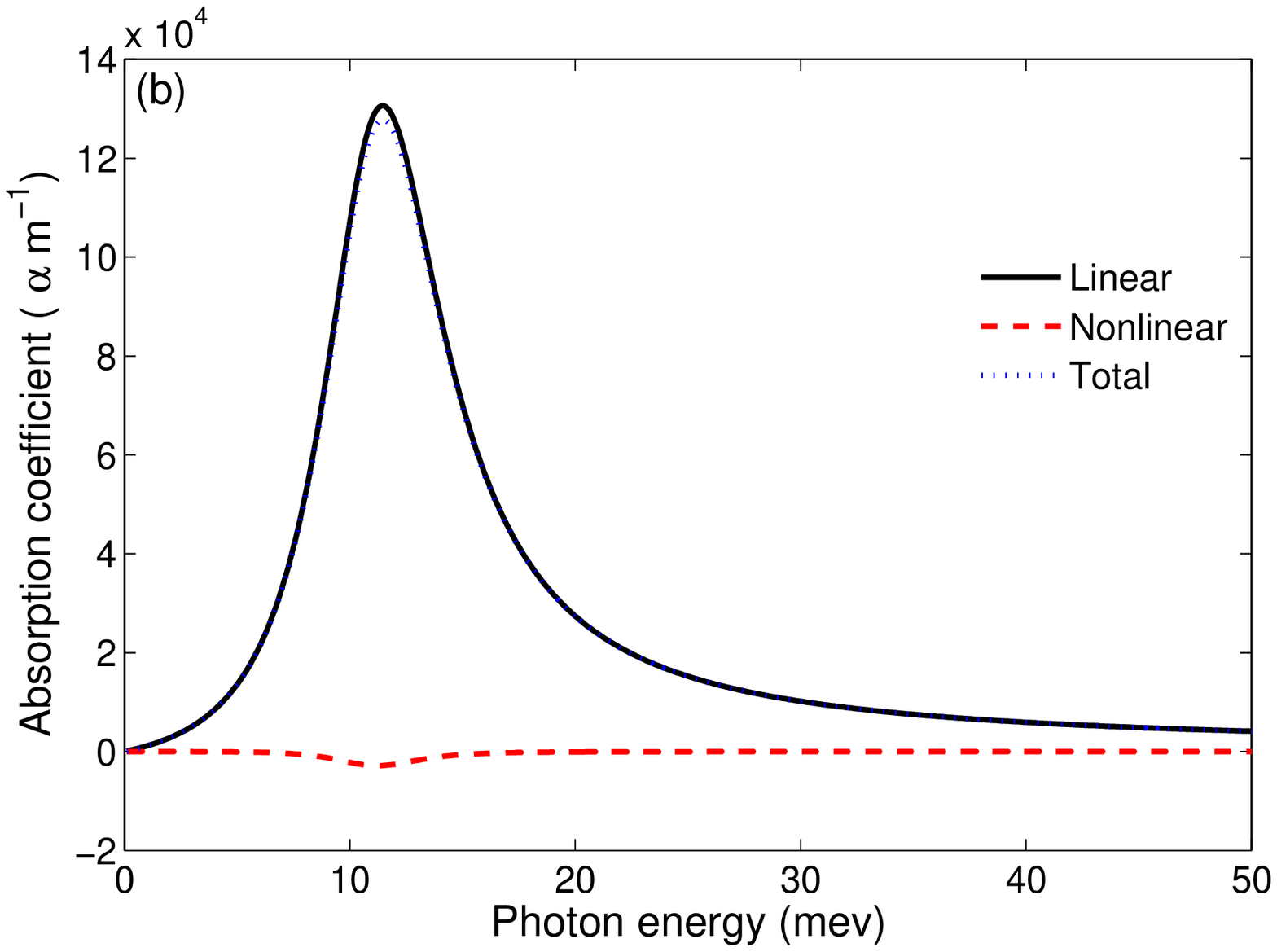} }
 \caption{The linear, the nonlinear and the total ACs as a function of incident photon energy in the same condition $\{I=20MW/m^{2}$, $b-a=0.5a_{0}$, $V_{1}=2Ry$ and $V_{2}=5Ry\}$ for (a) MLQAD (b) MLQD.}
 \label{f2}\end{figure*}

Figs.~\ref{f3} (a) and (b) show the linear, nonlinear and total absorption coefficients as a function
of the photon energy for MLQAD and MLQD respectively. The solid curves represent linear ACs where as in (22) are independent
of intensity, the dashed (dotted) curves display the nonlinear (total) ACs curves. It is observed that,as the optical intensity increases, the total AC
 decreases for both MLQD and MLQAD. This is because the nonlinear absorption (which is negative) enhances with an
increase in intensity. In Fig.~\ref{f3}(a) that shows the behavior of MLQAD, for total ACs curves, from up, the intensity value goes
from $I=5MW/m^{2}$ to $30MW/m^{2} $ at fixed incremental steps of $5MW/m^{2}$.
If this range of intensity is used for the MLQD, one can see that the MLQD has no significant changes in the nonlinear AC and therefore in
 the total AC (Fig.~\ref{f2}(b) is the special case of $I=20MW/m^{2}$ that the MLQD has no significant nonlinear AC) we used  the
  range $100\leq I \leq 600$ with a the fixed incremental steps of $100MW/m^{2}$ as the Fig.~\ref{f3}(b) shows it.
  In this case, at sufficiently  high intensity, the nonlinear term causes a collapse at the
center of the total AC peak that leads to the splitting of the AC peak into two peaks.
\begin{figure*}
\mbox{\includegraphics[ height=2.7in,width=3.5in]{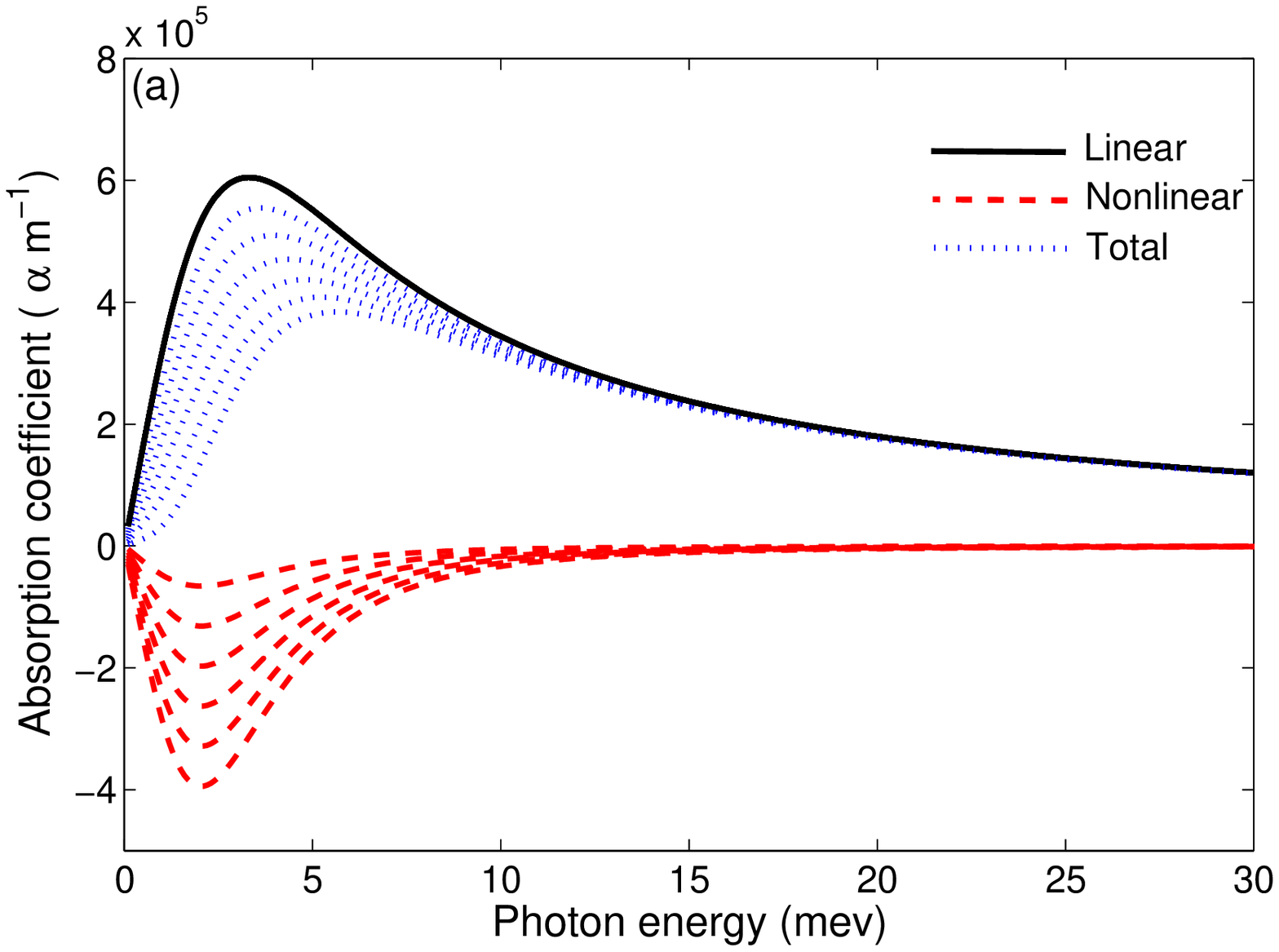}\quad
\includegraphics[ height=2.7in,width=3.5in]{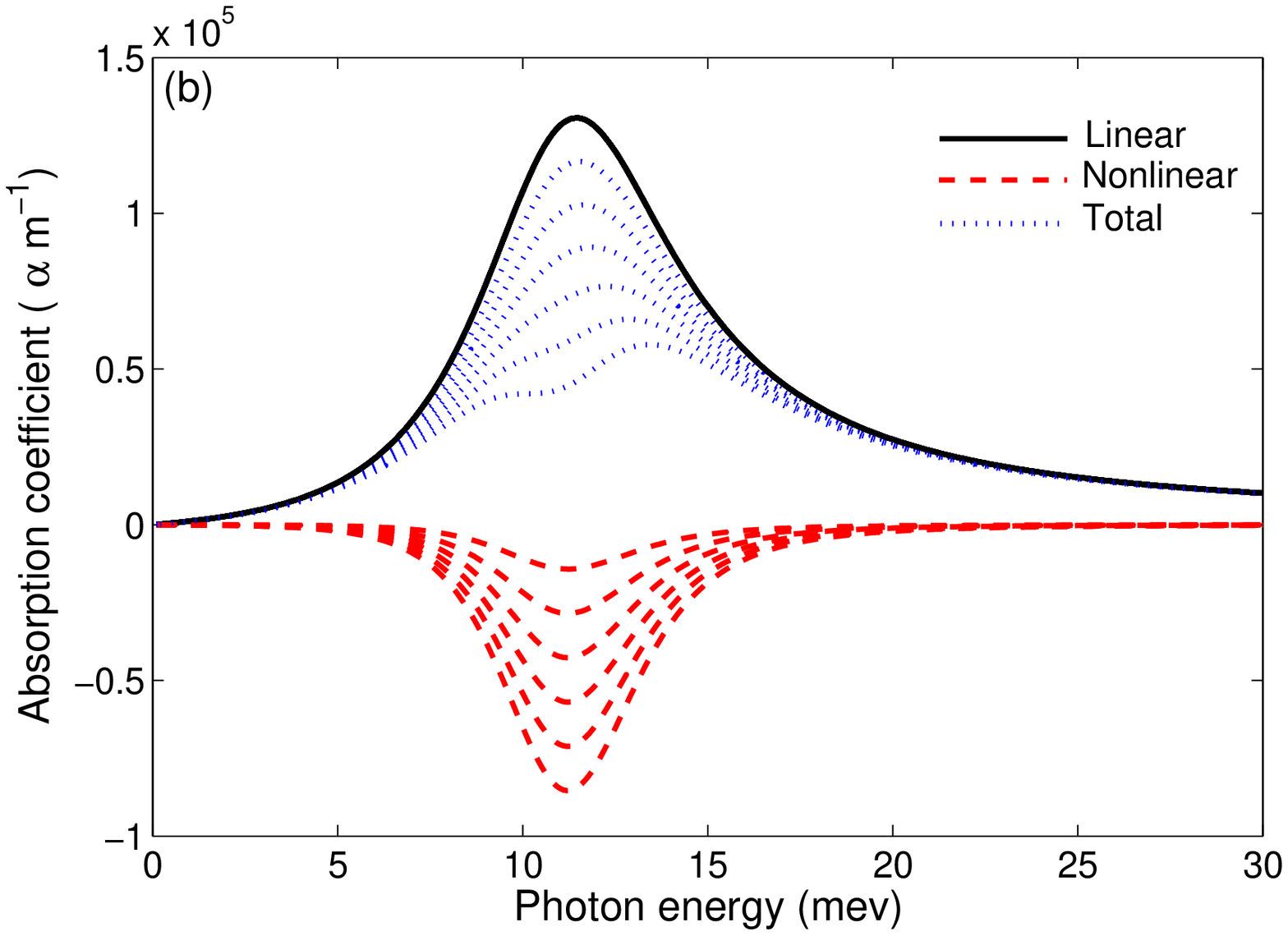} }
 \caption{The linear, the nonlinear and the total ACs as a function of incident photon energy with the fixed parameters $\{a=2a_{0}, b-a=2.5a_{0}, V_{1}=2Ry$ and $V_{2}=5Ry\}$ and six different intensities for (a) MLQAD that from up, the intensity value goes from $I=5MW/m^{2}$ to $30MW/m^{2} $ at fixed increments of $5MW/m^{2}$ (b) MLQD that the range of intensity is $100\leq I \leq 600$ with the fixed incremental steps of $100MW/m^{2}$.}
 \label{f3}\end{figure*}
\begin{figure*}
\mbox{\includegraphics[ height=2.7in,width=3.5in]{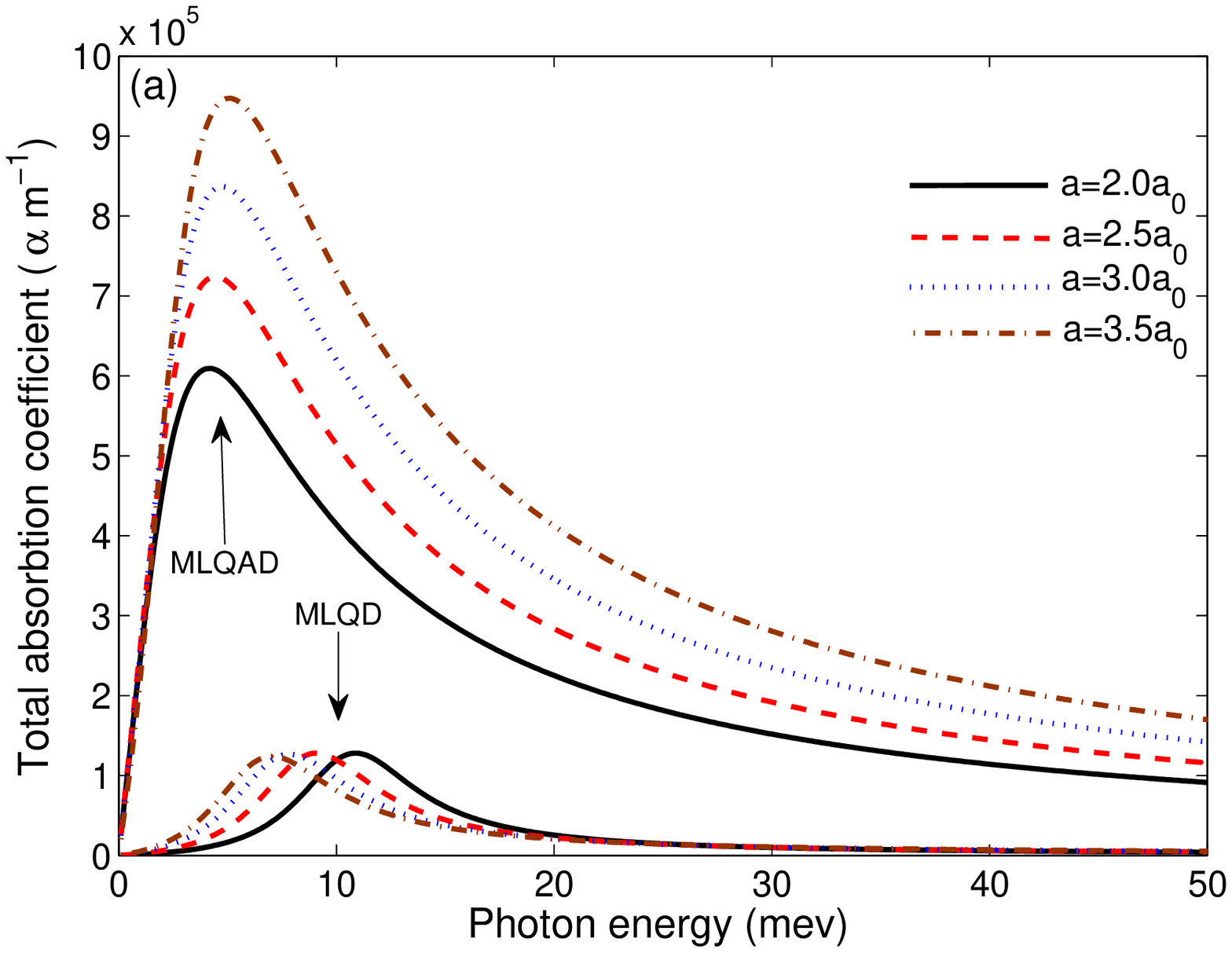}\quad
\includegraphics[ height=2.7in,width=3.5in]{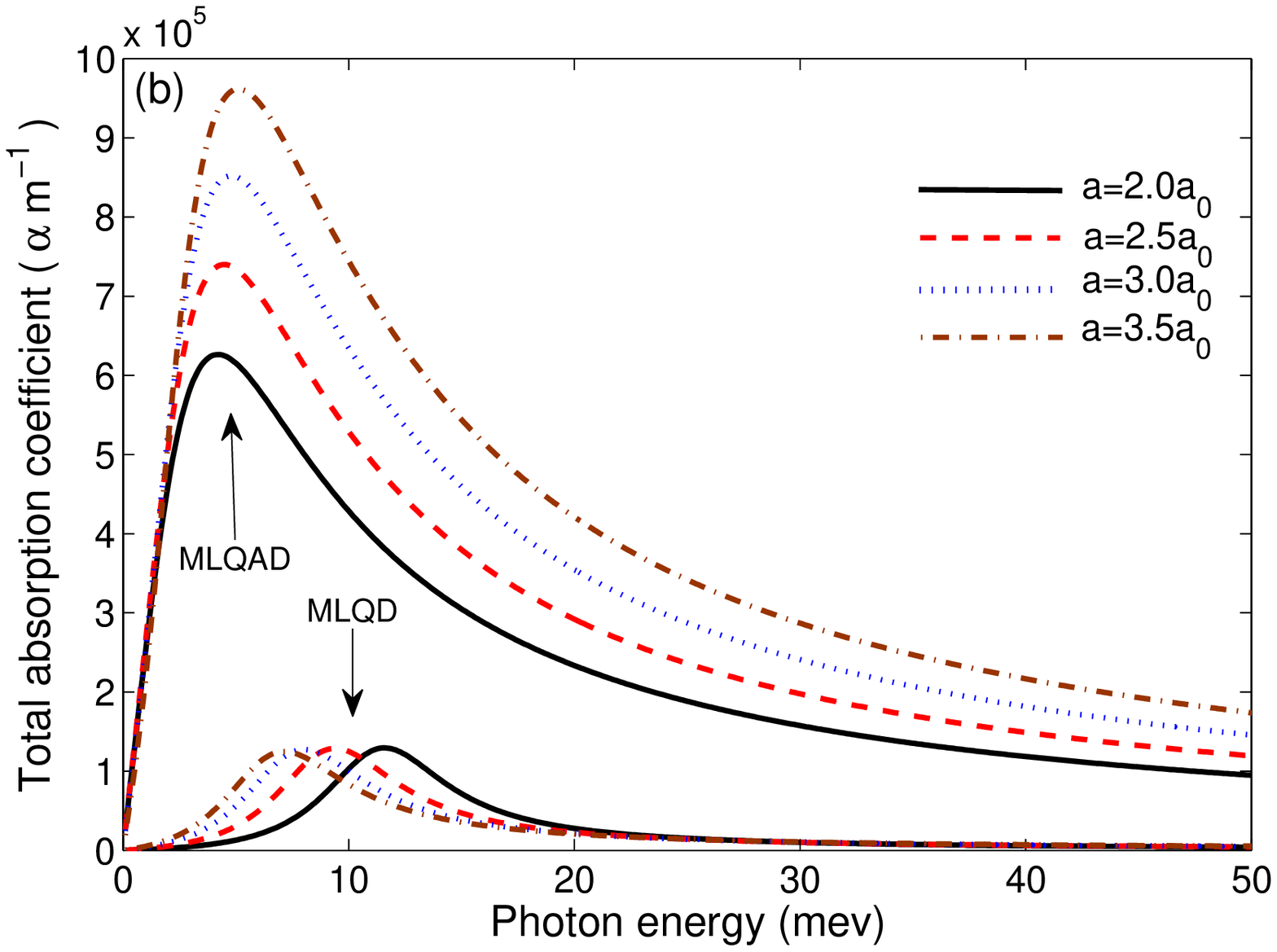} }
 \caption{The total AC of MLQD and MLQAD as a function of incident photon energy with the fixed shell thickness $b-a=1a_{0}$ and four different core radii for (a) $V_{1}=2Ry$ and $V_{2}=5Ry$ (b) $V_{1}=2Ry $ and $V_{2}=\infty Ry$.}
 \label{f4}\end{figure*}

In Fig.~\ref{f4}(a), the total AC of both the MLQAD and MLQD are plotted when the shell thickness and CPs are fixed
as $b-a=1a_{0}$ and ($V_{1}=2Ry, V_{2}=5Ry$) whiles a core radius has a four different values as $a=2, 2.5, 3, 3.5 a_{0}$.
The graph shows that in lower incident photon energies, the total AC is mainly happened in MLQAD case and by increasing the
amount of core radius, the total AC peak heights becomes larger and slightly shift toward larger incident photon energies.
In the case of the MQAD, by increasing the core radius value, the peak heights corresponding to MLQD has no significant changes,
but move toward the smaller incident photon energies. Figs.~\ref{f4}(a) and (b) are plotted in the same condition, however
their difference is in their $V_{2}$ values, where $V_{2}=5Ry$ in Fig.~\ref{f4}(a) and $V_{2}=\infty$ in Fig.~\ref{f4}(b). Regarding these
two figures, no considerable changes are observed.  \\

The total AC curves  of the MLQAD and MLQD for four different shell thickness $b-a=1, 0.5, 0.25, 0.02 a_{0}$, as the function of photon
energy are plotted in Figs.~\ref{f5}. In Fig.~\ref{f5}(a) the selected CPs and core radius are $V_{1}=2Ry, V_{2}=5Ry$ and $a=2.0a_{0}$
that by increasing the shell thickness, the peak of total AC curves corresponding to MLQAD moves to larger total AC values without
considerable changes in incident photon energy, however the curves related to MLQD shift to smaller energy regions without considerable
changes in peak heights. Fig.~\ref{f5}(b) shows the behaviour of the MLQAD and MLQD with the same conditions as Fig.~\ref{f5}(a)
with new CPs $(V_{1}=2Ry, V_{2}=\infty Ry)$.  By comparing the mutual corresponding curves for example dash-dot curves in
Figs.~\ref{f5} (a) and (b), it is observed that in the biggest shell thickness one can ignore the differences between the total
AC curves but in the smaller shell thicknesses, the curves are more affected by  changes in CPs.

Generally, from Figs.~\ref{f4} and ~\ref{f5}
it is found that under the same conditions,
the total AC in MLQAD is greater than MLQD and also by increasing the shell thickness and core radius,
the peak height of MLQAD is more sensitive and becomes
significant larger, so we conclude that the utilization of MLQAD structures is more useful.
\begin{figure*}
\mbox{\includegraphics[ height=2.7in,width=3.5in]{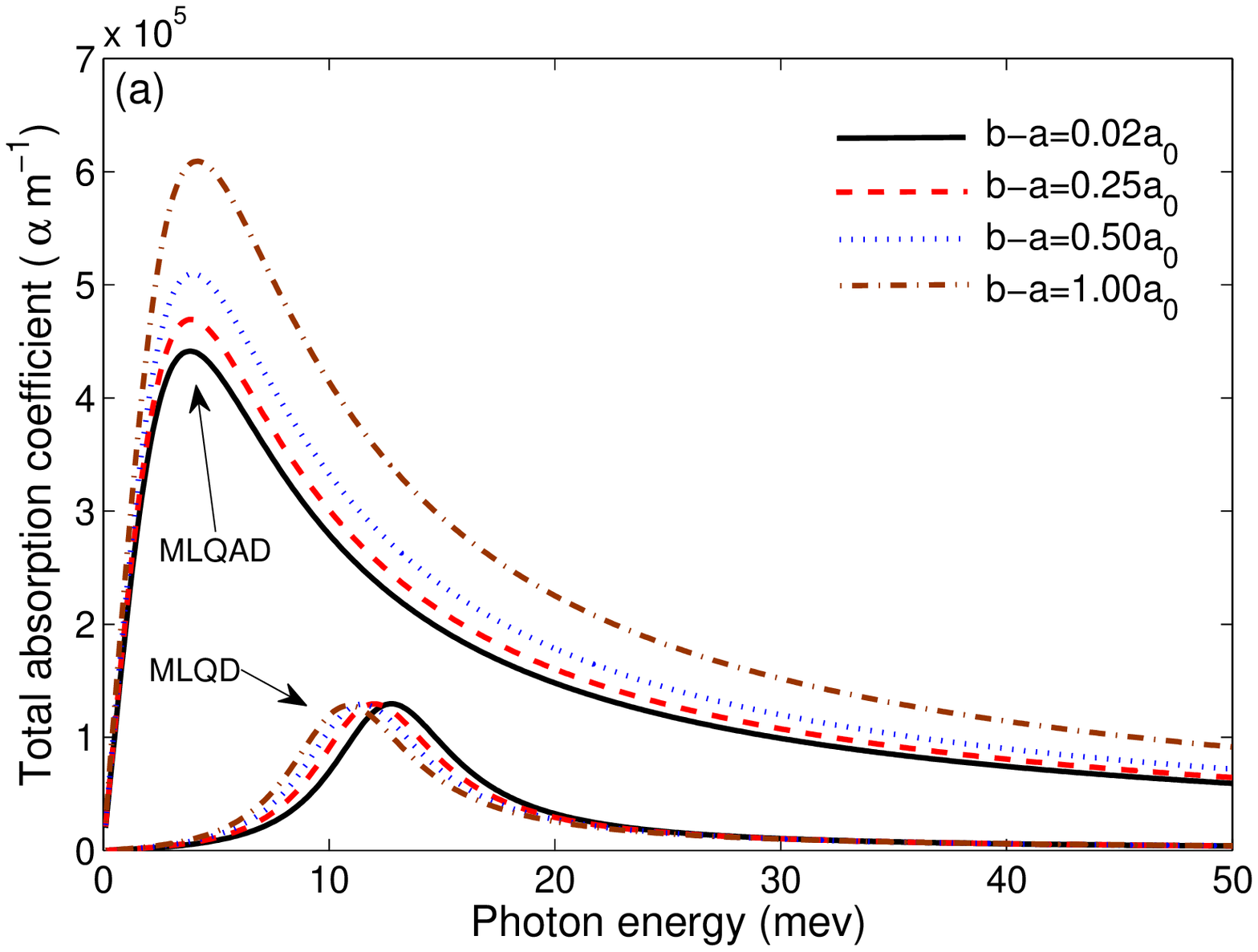}\quad
\includegraphics[ height=2.7in,width=3.5in]{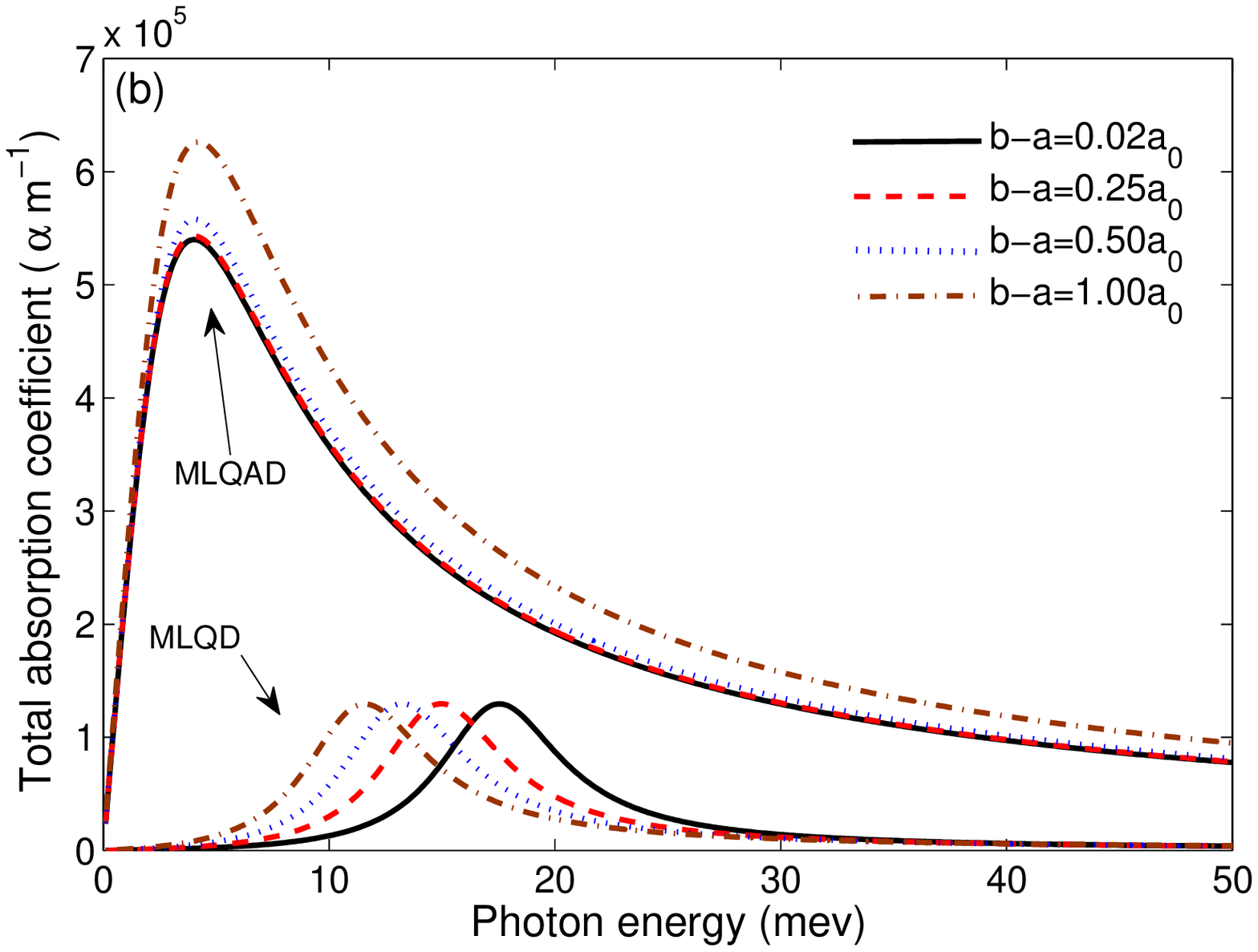} }
 \caption{The total AC of MLQD and MLQAD as a function of incident photon energy with $a=2a_{0}$ and four different shell thicknesses for (a) $V_{1}=2Ry$ and $V_{2}=5Ry$ (b) $V_{1}=2Ry $ and $V_{2}=\infty Ry$.}
 \label{f5}\end{figure*}

\section{Conclusion}
By studying the case of multi-layered spherical nano-systems with a hydrogenic impurity, the linear,  third-order nonlinear
and total optical ACs have been calculated. As our results indicates, in the case of the MLQAD the absorption curves abruptly increase
then asymptotically go to zero but in the case of MLQD the absorption curves have symmetrical shapes. Furthermore, the range of
variation of intensity that leads to significant changes in total AC curves is different for MLQAD and MLQD.
It is observed that for the fixed shell thickness, by increasing the amount of core radius, the total AC peak heights become larger
and slightly shift toward larger incident photon energies for MLQAD but the peak heights corresponding to MLQD have no
significant changes but move toward the smaller incident photon energies. Also, in this case the change in CPs leads to no considerable
 changes in total AC curves. Moreover, for a fixed core radius, by increasing the shell thickness, the peak of total AC curves
  corresponding to MLQAD, moves to larger values without considerable changes in the incident photon energy but the
curves related to MLQD shift to smaller energy regions without considerable changes in
peak heights. Finally, it is found that in the latter case, by decreasing the shell thicknesses,
 the total AC curves are considerably affected by CPs values.
\begin{acknowledgements}
The authors wish to thank H. Mosavi and M. Rahimi for a number of useful comments and suggestions.
\end{acknowledgements}
\end{document}